\documentclass[twocolumn,showpacs,aps,prl,superscriptaddress]{revtex4}

\usepackage{graphicx}
\usepackage{dcolumn}
\usepackage{bm}

\begin{document}

\title{Strange Quark Contributions to Parity-Violating Asymmetries in the Forward G0 Electron-Proton Scattering Experiment
}

\author{D.~S.~Armstrong}
\affiliation{Department of Physics, College of William and Mary, Williamsburg, VA 23187 USA} 

\author{J.~Arvieux}
\affiliation{Institut de Physique Nucl\'eaire d'Orsay, Universit\'e Paris-Sud, F-91406 Orsay Cedex FRANCE}

\author{R.~Asaturyan}
\affiliation{Yerevan Physics Institute, Yerevan 375036 ARMENIA}

\author{T.~Averett}
\affiliation{Department of Physics, College of William and Mary, Williamsburg, VA 23187 USA} 

\author{S.~L.~Bailey}
\affiliation{Department of Physics, College of William and Mary, Williamsburg, VA 23187 USA} 

\author{G.~Batigne}
\affiliation{Laboratoire de Physique Subatomique et de Cosmologie, F-38026 Grenoble Cedex FRANCE}

\author{D.~H.~Beck}
\affiliation{Loomis Laboratory of Physics, University of Illinois, Urbana, IL 61801 USA}

\author{E.~J.~Beise}
\affiliation{Physics Department, University of Maryland, College Park, MD 20472 USA}

\author{J.~Benesch}
\affiliation{Thomas Jefferson National Accelerator Facility, Newport News, VA 23606 USA}

\author{L.~Bimbot}
\affiliation{Institut de Physique Nucl\'eaire d'Orsay, Universit\'e Paris-Sud, F-91406 Orsay Cedex FRANCE}

% ...

\collaboration{G0 Collaboration}

\noaffiliation

\date{\today}

\begin{abstract}
We have measured parity-violating asymmetries in elastic electron-proton scattering over the range of momentum transfers $0.12 \le Q^2 \le 1.0$ GeV$^2$.  These asymmetries, arising from interference of the electromagnetic and neutral weak interactions, are sensitive to strange quark contributions to the currents of the proton.  The measurements were made at JLab using a toroidal spectrometer to detect the recoiling protons from a liquid hydrogen target.  The results indicate non-zero, $Q^2$ dependent, strange quark contributions and provide new information beyond that obtained in previous experiments.
\end{abstract}


\maketitle

At short distance scales, bound systems of quarks have relatively simple properties and QCD is successfully described by perturbation theory.  On size scales similar to that of the bound state itself, $\sim 1$ fm, however, the QCD coupling constant is large and the effects of the color fields cannot yet be calculated accurately,
even in lattice QCD.  In addition to valence quarks, e.g., $uud$ for the proton, there is a sea of gluons and $q \bar q$ pairs that plays an important role at these distance scales.  From a series of experiments measuring the neutral weak scattering of electrons from protons and neutrons, we can extract the contributions of strange quarks to the ground state charge and magnetization distributions (e.g., magnetic moment) of the nucleon.  These strange quark contributions must originate in fluctuations of gluons to $s \bar s$ pairs because there are no strange valence quarks in the nucleon.   
There have been numerous estimates of strange quark contributions to nucleon properties within various phenomenological models and also in state-of-the-art lattice calculations~\cite{beckholstein,leinweber02}; many focus on the contribution to the magnetic moment.
In this paper, we report on a new measurement sensitive to strange quark contributions over a range of distance scales.

Separation of the strange quark contributions to nucleon currents in the context of the neutral weak interaction dates back to Cahn and Gilman~\cite{cahn78a} and was developed by Kaplan and Manohar~\cite{kaplan88a}.  Because the coupling of both photons and $Z$ bosons to point-like quarks is well defined, it is possible, by comparing the corresponding currents, to separate the contributions of the various flavors~\cite{mckeown89a,beck89a,beckmckeown}.  The charge and magnetic form factors of the proton can be written ($i=\gamma,Z$) 
\begin{equation}
\label{eqn:protonff}
G_{E,M}^{p,i}= e^{i,u} G_{E,M}^u + e^{i,d} \left( G_{E,M}^d + G_{E,M}^s \right), \nonumber
\end{equation}
neglecting the very small contribution from heavier flavors.  For the ordinary electromagnetic form factors the charges are $e^\gamma = +2/3$, $-1/3$ for $u$ and $d/s$ quarks, respectively.  Assuming that the proton and neutron are related by a simple exchange of $u$ and $d$ quarks~\cite{miller98} (and the corresponding anti-quarks), the ordinary neutron form factors can be written in terms of these same contributions
\begin{equation}
G_{E,M}^{n,\gamma}= \frac{2}{3} G_{E,M}^d - \frac{1}{3} \left( G_{E,M}^u + G_{E,M}^s \right). \nonumber
\end{equation}
A complete separation of the $G_{E,M}^q$, and, in particular, isolation of $G_{E,M}^s$, requires a third combination.  
In this paper, new measurements of the weak interaction form factors of the proton are presented which allow us to determine the strange quark contributions. These form factors are written (Eqn.\ \ref{eqn:protonff}) in terms of the weak charges, $e^Z = 1-8/3\ \hbox{sin}^2\theta_W$, $-1+4/3\ \hbox{sin}^2\theta_W$ for the $u$ and $d/s$ quarks, respectively, where $\theta_W$ is the weak mixing angle.

In order to isolate the small contribution to elastic electron-proton scattering from the neutral weak current, we measure the parity-violating asymmetry for longitudinally polarized ($R$ and $L$) electrons~\cite{beckmckeown}
\begin{widetext}
\begin{equation}
A = {d\sigma_R - d\sigma_L \over d\sigma_R + d\sigma_L}
= - \frac{G_F Q^2}{4 \sqrt{2} \pi \alpha}
{{
\varepsilon G^{\gamma}_{{E}} G^{Z}_
{{E}} + \tau G^{\gamma}_{{M}}
G^{Z}_{{M}} - (1-4 \sin^2 \theta_W ) 
\varepsilon^{\prime} G^{\gamma}_{{M}} G^{e}_{A}}   \over
{\cal D}}
\label{eqn:asym}
\end{equation}
\end{widetext}
where
\begin{eqnarray}
\tau = {{Q^2} \over {4 M_p^2}}, \ 
\varepsilon = \left (1 + 2(1 + \tau)\tan^2{\theta \over 2} \right )^{-1}, \nonumber \\
{\cal D} = \varepsilon (G^\gamma_{E})^2 + \tau (G^\gamma_{M})^2, \hbox{ and }
\varepsilon^{\prime} = \sqrt{\tau (1+\tau) (1- \varepsilon^2)}, \nonumber
\end{eqnarray}
$Q^2$ is the squared four-momentum transfer ($Q^2 > 0$), $G_F$ and $\alpha$ the usual weak and electromagnetic couplings, $M_p$ the proton mass and
$\theta$ the laboratory electron scattering angle.  The three new form factors in this asymmetry, $G_E^Z$, $G_M^Z$ and $G_A^e$ may be separated by measuring elastic scattering from the proton at forward and backward angles, and quasi-elastic scattering from the deuteron at backward angles~\cite{beckmckeown}.  

The G0 experiment~\cite{G0equip} was performed in Hall C at Jefferson Lab.  We used a 40 $\mu A$ polarized electron beam with an energy of $3.031 \pm 0.001$ GeV over the measurement period of 700 h.
It was generated with a strained GaAs polarized source~\cite{poelker00a} with 32 ns pulse timing (rather than the standard 2 ns) to allow for time-of-flight (t.o.f.) measurements.  The average beam polarization, measured with a M\o ller polarimeter~\cite{hauger99} in interleaved runs, was $73.7 \pm 1.0\%$.
Helicity-correlated current and position changes were corrected with active feedback to levels of about 0.3 parts-per-million (ppm) and 10 nm, respectively.  Corrections to the measured asymmetry were applied via linear regression for residual helicity-correlated beam current, position, angle and energy variations and amounted to a negligible total of 0.02 ppm; the largest correction was 0.01 ppm for helicity-correlated current variation.
We made one further correction of, on average, $+0.71 \pm 0.14$ ppm to the asymmetries in all detectors ($\sim 5$\% variation from detector to detector).  It was associated with a small ($\sim 10^{-3}$) fraction of the beam current with a 2 ns structure (``leakage beam'': tails of beams from other operating halls) and a large charge asymmetry ($\sim 570$ ppm); it was measured in otherwise `forbidden' regions of the t.o.f.\ spectra.

\begin{figure}
\resizebox{20.5pc}{!}{\includegraphics{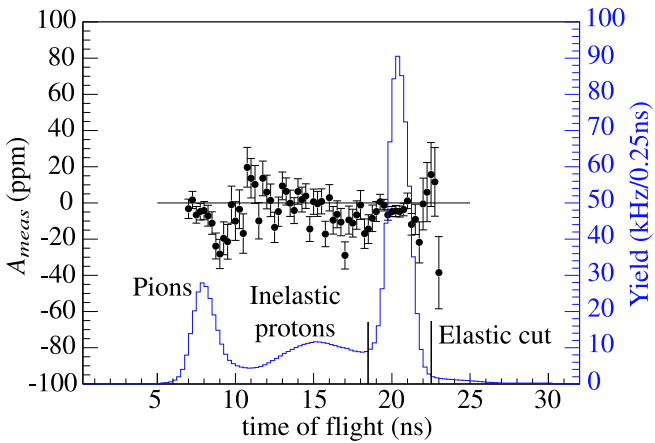}}
\caption{\label{fig:tof} Example of the raw asymmetry, $A_{meas}$, (data points) and yield (histogram) as a function of t.o.f.\ for detector 8.}
\end{figure}

The polarized electrons scattered from a 20 cm liquid hydrogen target~\cite{G0targ}; the recoiling elastic protons were detected to allow simultaneous measurement of the wide range of momentum transfer, $0.12 \le Q^2 \le 1.0$ GeV$^2$.  This was effected using a novel toroidal spectrometer designed to measure the entire range with a single field setting and with precision comparable to previous experiments.  The spectrometer included an eight-coil superconducting magnet and eight sets of scintillator detectors.  Each set consisted of 16 scintillator pairs used in coincidence to cover the range of momentum transfers (smallest detector number corresponding to the lowest momentum transfer).  Because of the correlation between the momentum and scattering angle of the elastic protons (higher momentum corresponds to more forward proton scattering angles), detector 15 covered the range of momentum transfers between 0.44 and 0.88 GeV$^2$ which we divided into three t.o.f.\ bins with average momentum transfers of 0.51, 0.63 and 0.79 GeV$^2$.  For the same reason detector 14 had two elastic peaks separated in t.o.f.\ with momentum transfers of 0.41 and 1.0 GeV$^2$; detector 16, used to determine backgrounds, had no elastic acceptance.  Custom time-encoding electronics
sorted detector events by t.o.f.; elastic protons arrived about 20 ns after the passage of the electron bunch through the target (see Fig.\ \ref{fig:tof}).  The spectrometer field integral and ultimately the $Q^2$ calibration ($\Delta Q^2/Q^2 = 1$\%) was fine-tuned using the measured t.o.f.\ difference between pions and elastic protons for each detector.
All rates were corrected for dead-times of $10-15\%$ on the basis of the measured yield dependence on beam current; the corresponding uncertainty in the asymmetry is $\sim 0.05$ ppm.  Standard radiative corrections~\cite{afanasev01a} in the range of 1 - 3\%, determined by comparing simulations with and without radiation, were also applied to the asymmetries. 
Lastly, there is an uncertainty of 0.01 ppm due to a small component of transverse polarization in the beam.
 
As shown in Fig.\ \ref{fig:tof}, a background extends on both sides of the elastic proton peak at a t.o.f.\ $\sim 20$ ns.  This background is essentially all protons (as determined from energy loss measurements in a sampled data set): quasi-elastic protons from the aluminum target windows and inelastic protons from both the hydrogen and the aluminum.  
The measured asymmetry has two components
\begin{equation}
A_{meas}=\left( 1-f \right) A_{el}+ f A_{back} \nonumber
\end{equation}
where $A_{el}$ is the raw elastic asymmetry and $f$ is the background fraction; in the actual analysis t.o.f.\ fits to the yield and asymmetry in the region of the elastic peak are used.  The yield is typically modeled with a Gaussian elastic peak and a polynomial background.  The asymmetry model comprises a quadratic background and a constant for the elastic.  For higher numbered detectors the background asymmetry is positive.  In particular, for detector 15 the background asymmetry has a maximum value of about 45 ppm in the region of the elastic peak.  As substantiated by a Monte Carlo simulation, this positive asymmetry is caused by a small number of $\Lambda$ and $\Sigma$ weak-decay protons scattered inside the spectrometer magnet.  The smooth variation of the region of positive asymmetries is tracked from detectors 12-14 through to detector 16; the background asymmetry for the large acceptance of detector 15 is then corrected by interpolating these measured background asymmetries.  As a check, the same fitting procedure described above is also used for detector 15 and gives consistent results.

The elastic asymmetries for the experiment, $A_{phys}$ ($A_{el}$ corrected for all effects described earlier) are presented in Table \ref{tab:asymmetries}.
\begin{table}
\caption{\label{tab:asymmetries} Asymmetries and uncertainties measured in the present experiment~\cite{web}.  The contributions to the systematic uncertainties are summarized in Table \ref{tab:unc}.}
\begin{ruledtabular}

\begin{tabular}{lcccc|cc}
$Q^2$&$A_{phys}$&$\Delta A_{stat}$&$\Delta A_{pt-pt}$&$\Delta A_{glob}$&$f$&$A_{meas}$\\
(GeV$^2$)&(ppm)&(ppm)&(ppm)&(ppm) & & (ppm)\\
\hline
0.122&	-1.51&	0.44&	0.22&	0.18&	0.061&	-1.38\\
0.128&	-0.97&	0.41&	0.20&	0.17&	0.084&	-1.07\\
0.136&	-1.30&	0.42&	0.17&	0.17&	0.085&	-1.34\\
0.144&	-2.71&	0.43&	0.18&	0.18&	0.077&	-2.67\\
0.153&	-2.22&	0.43&	0.28&	0.21&	0.096&	-2.46\\
0.164&	-2.88&	0.43&	0.32&	0.23&	0.100&	-3.13\\
0.177&	-3.95&	0.43&	0.25&	0.20&	0.110&	-4.47\\
0.192&	-3.85&	0.48&	0.22&	0.19&	0.110&	-5.01\\
0.210&	-4.68&	0.47&	0.26&	0.21&	0.116&	-5.73\\
0.232&	-5.27&	0.51&	0.30&	0.23&	0.136&	-6.08\\
0.262&	-5.26&	0.52&	0.11&	0.17&	0.154&	-5.55\\
0.299&	-7.72&	0.60&	0.53&	0.35&	0.174&	-5.40\\
0.344&	-8.40&	0.68&	0.85&	0.52&	0.182&	-3.65\\
0.410&	-10.25&	0.67&	0.89&	0.55&	0.180&	-1.70\\
0.511&	-16.81&	0.89&	1.48&	1.50&	0.190&	-5.80\\
0.631&	-19.96&	1.11&	1.28&	1.31&	0.20&	-9.74\\
0.788&	-30.8&	1.9&	2.6&	2.59&	0.40&	-12.66\\
0.997&	-37.9&	7.2&	9.0&	0.52&	0.78&	4.21\\
\end{tabular}
\end{ruledtabular}
\end{table}
\begin{table}
\caption{\label{tab:unc} Systematic uncertainties for measured asymmetries.  The first six uncertainties are global, deadtime is point-to-point and the background is a combination (see text).}
\begin{ruledtabular}
\begin{tabular}{lc}
Source&Uncertainty\\
\hline
Helicity-correlated beam parameters &  0.01 ppm\\
Leakage beam &  0.14 ppm\\
Beam polarization &  1.0\%\\
Ordinary radiative corrections & 0.3\% \\
Transverse polarization & 0.01 ppm \\
$Q^2$	& 1\%\\
Background correction& 0.2 - 9 ppm \\
Deadtime &  0.05 ppm\\
\end{tabular}
\end{ruledtabular}
\end{table}
The statistical uncertainties include those from the measured and the background asymmetries.  The systematic uncertainties (Table \ref{tab:unc}) are dominated by those from the background correction.  This uncertainty is estimated from the range of elastic asymmetries generated from a variety of different background yield and asymmetry models.  These models are bounded by the measured slopes of background yields and asymmetries on either side of the elastic peak and varied continuously between these limits.  The uncertainty in the background asymmetry for detector 15 is conservatively taken to be the difference between interpolated background asymmetries in successive detectors as described above.  We have also estimated the global and point-to-point contributions to these uncertainties from the extent to which a change in, e.g., the background asymmetry functional form, consistently changes the asymmetries in all the affected detectors.  

\begin{figure}
\resizebox{20pc}{!}{\includegraphics{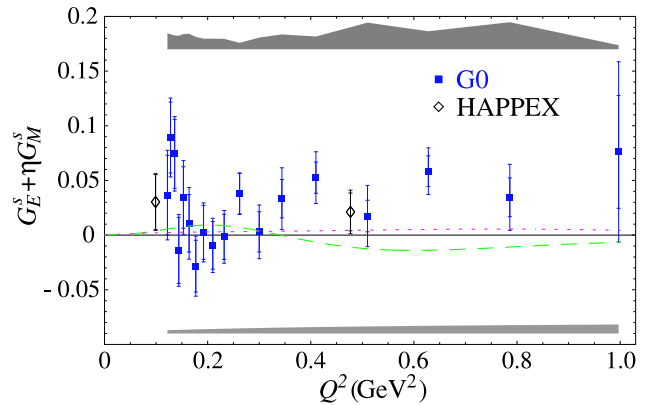}}
\caption{\label{fig:geshgms}The combination $G_E^s + \eta G_M^s$ for the present measurement.
The gray bands indicate systematic uncertainties (to be added in quadrature); the lines correspond to different electromagnetic nucleon form factor models (see text).}
\end{figure}

The results of the experiment are shown as a function of momentum transfer in Fig.\ \ref{fig:geshgms}.  The quantity
\begin{equation}
G_E^s + \eta  G_M^s = 
\frac{4 \sqrt{2} \pi \alpha}{G_F Q^2}
\frac
{\cal D}
{\varepsilon G^\gamma_{E}}
\left(A_{phys} - A_{NVS}\right),
\label{eqn:gesgms}
\end{equation}
(where $\eta \left( Q^2 \right) = \tau G_M^p / \varepsilon G_E^p$)
is determined from the difference between the experimental asymmetry and the ``no-vector-strange'' asymmetry, $A_{NVS}$. $A_{NVS}$ is calculated from Eqn.\ \ref{eqn:asym} with $G_E^s = G_M^s = 0$ for all values of $Q^2$, and using the electromagnetic form factors of Kelly~\cite{kelly04}.  Also shown is the excellent agreement with the HAPPEX measurements~\cite{happexI,happexII} made at nearly the same kinematic points (with small corrections to the asymmetries, $<0.2$ ppm,  to adjust them to the G0 beam energy).  The error bars include the statistical uncertainty (inner) and statistical plus point-to-point systematic uncertainties added in quadrature (outer).  The error bands represent, for the G0 experiment, the global systematic uncertainties: from the measurement (upper) and from the uncertainties in the quantities entering $A_{NVS}$ (lower).  These quantities are: 
the calculated value of the axial-vector form factor normalization~\cite{zhu00a} (differing from $g_A/g_V$ by electroweak radiative corrections), the same dipole momentum transfer dependence for $G_A^e(Q^2)$ as is deduced for $G_A(Q^2)$~\cite{bernard01a}, 
the axial vector strangeness contribution $\Delta s$~\cite{leader03}, and the electroweak radiative corrections~\cite{musolf93a}.  The sensitivity of the result to electromagnetic form factors is shown separately by the lines on the plot. For the alternative form factor parameterizations of Friedrich and Walcher (FW)~\cite{fw03} (dashed) and the combination (dotted): Arrington ``Rosenbluth''~\cite{arrington04} - proton, and Kelly~\cite{kelly04} - neutron, the effective $A_{NVS}$ is shown (e.g., for the FW parameterization, the value of $G_E^s + \eta G_M^s$ at $Q^2 = 0.63$ GeV$^2$ increases from 0.059 to 0.072).  Alternately, the uncertainties in the Kelly form factor fits would increase the width of the uncertainty band (lower) for $A_{NVS}$ at each $Q^2$ by about 25\% if included there.  

The $G_E^s + \eta G_M^s$ data shown in Fig.\ \ref{fig:geshgms} have a systematic and intriguing $Q^2$ dependence.  For reference we note that $G_E^s + \eta  G_M^s=0$ at $Q^2 = 0$ and that $\eta \sim 0.94Q^2$ (Kelly form factors) for our kinematics.  First, to characterize our result with a single number, we tested the hypothesis $G_E^s + \eta  G_M^s=0$ by generating randomized data sets with this constraint, distributed according to our statistical and systematic uncertainties (including correlated uncertainties).  The fraction of these with $\chi^2$ larger than that of our data set was 11\%, so we conclude that the non-strange hypothesis is disfavored with 89\% confidence.
More important is the $Q^2$ dependence of the data. The initial rise from zero to about 0.05 is consistent with the finding that $G_M^s(Q^2 = 0.1 \hbox{\ GeV$^2$}) \sim +0.5$ from the SAMPLE~\cite{sample1}, PVA4~\cite{pva4II} and HAPPEX~\cite{happexII} measurements. Because $\eta$ increases linearly throughout, the apparent decline of the data in the intermediate region up to $Q^2 \sim 0.3$ indicates that $G_E^s$ may be {\it negative} in this range.  There is also some support for this conclusion from the combination of G0 and PVA4~\cite{pva4I} results at $Q^2 = 0.23$ GeV$^2$.  There is a significant trend, consistent with HAPPEX~\cite{happexI}, to positive values of $G_E^s + \eta  G_M^s$ at higher $Q^2$.  Experiments planned for Jefferson Lab, including G0 measurements at backward angles, and MAMI (Mainz) will provide precise separations of $G_E^s$ and $G_M^s$ over a range of $Q^2$ to address this situation.

In summary, we have measured forward angle parity-violating asymmetries in elastic electron-proton scattering over a range of momentum transfers from 0.12 to 1.0 GeV$^2$.  These asymmetries determine the neutral weak interaction analogs of the ordinary charge and magnetization form factors of the proton.  From the asymmetries we have determined combinations of the strange quark contributions to these form factors, $G_E^s + \eta G_M^s$, which, together with other experiments, indicate that both $G_M^s$ and $G_E^s$ are non-zero.

\begin{acknowledgments}
We gratefully acknowledge the strong technical contributions to this experiment from many groups: Caltech, Illinois, LPSC-Grenoble, IPN-Orsay, TRIUMF and particularly the Accelerator and Hall C groups at Jefferson Lab.  This work is supported in part by CNRS (France), DOE (U.S.), NSERC (Canada) and NSF (U.S.).
\end{acknowledgments}

\end{document}